\def\year{2019}\relax
\documentclass[letterpaper]{article} 
\usepackage{aaai19}  
\usepackage{times}  
\usepackage{helvet}  
\usepackage{courier}  
\usepackage{url}  
\usepackage{graphicx}  
\usepackage{amsfonts}
\usepackage{subfigure}
\usepackage{multirow}
\usepackage{makecell}
\usepackage{booktabs}
\begin{document}
%
\title{A novel metric for  community detection}
\author{
Ke-ke Shang\textsuperscript{\rm 1}\thanks{Ke-ke Shang is the corresponding author},
Michael Small\textsuperscript{\rm 2,\rm 3},
Yan Wang\textsuperscript{\rm 1},
Di Yin\textsuperscript{\rm 1},
Shu Li\textsuperscript{\rm 1},
\\
\textsuperscript{\rm 1} Computational Communication Collaboratory, Nanjing University, Nanjing, 210093, P.R. China\\
\textsuperscript{\rm 2} Complex Systems Group, Department of Mathematics and Statistics, The University of Western Australia, Crawley, Western Australia 6009, Australia\\
\textsuperscript{\rm 3} Mineral Resources, CSIRO, Kensington, WA, 6151, Australia\\
\ kekeshang@nju.edu.cn,~
\ keke.shang.1989@gmail.com
}

\maketitle
\begin{abstract}

Research into detection of dense communities has recently attracted increasing attention within network science, various metrics for detection of such communities have been proposed. The most popular metric -- {\it Modularity} -- is based on the so-called rule that the links within communities are denser than external links among communities, has become the default.
However, this default metric suffers from ambiguity, and worse, all augmentations of modularity and based on a narrow intuition of what it means to form a ``community''. We argue that in specific, but quite common systems, links within a community are not necessarily more common than links between communities. Instead we propose that the defining characteristic of a community is that links are more predictable within a community rather than between communities. In this paper, based on the effect of communities on link prediction, we propose a novel metric for the community detection based directly on this feature. We find that our metric is more robustness than traditional modularity. Consequently, we can achieve an evaluation of algorithm stability for the same detection algorithm in different networks. Our metric also can directly uncover the false community detection, and infer more statistical characteristics for detection algorithms.  

\end{abstract}

\section{Introduction}
\begin{figure}[htbp]
\centering
\includegraphics[width=\linewidth]{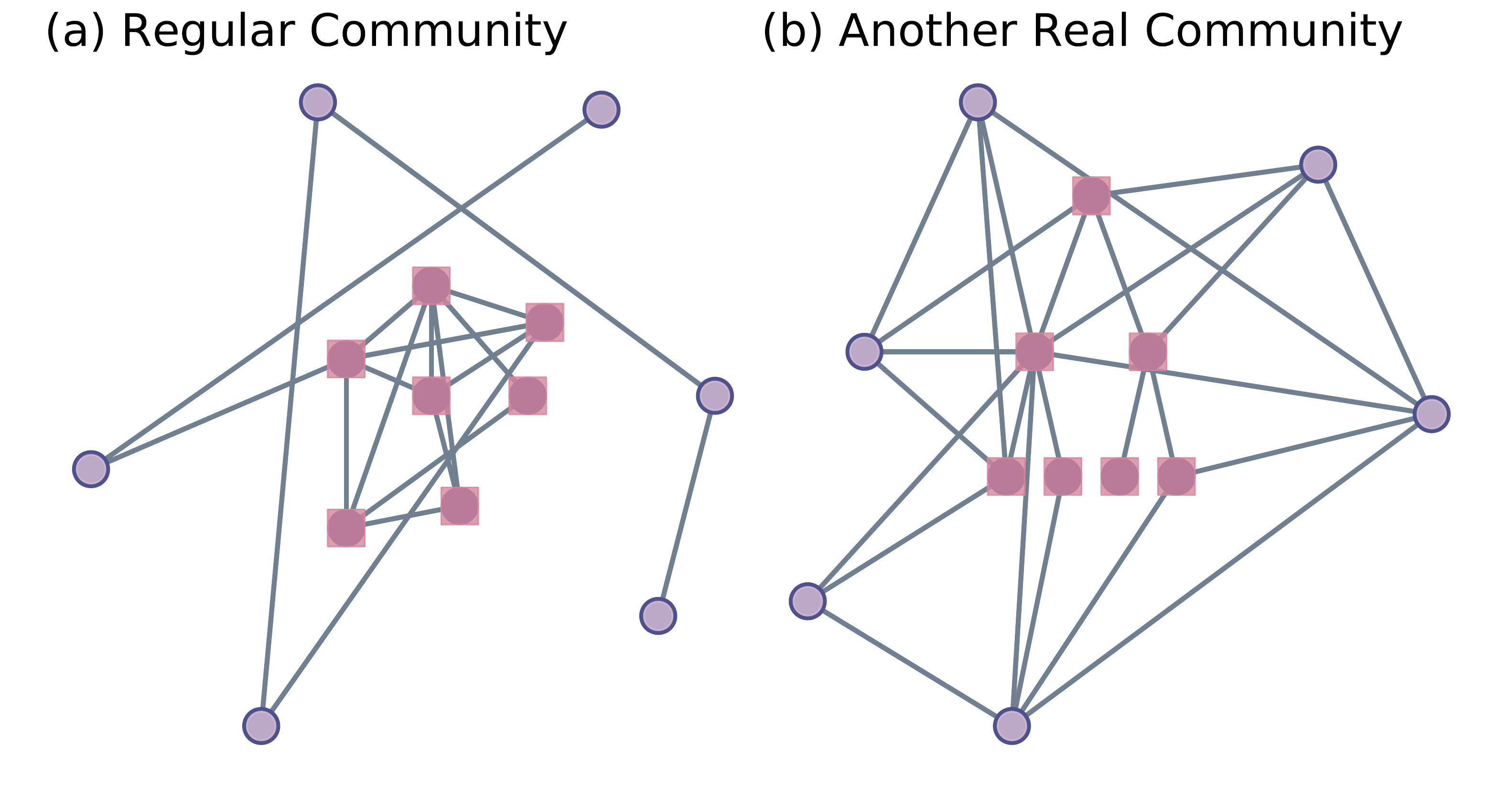}
\caption{Two different kinds of communities. The squares indicate  nodes within a single community. Conversely, the circles indicate nodes external to the community. Panel (a) depicts the traditional community which was been well-studied by previous studies. Panel (b) is an another real-world (``corporate'') community which may (for example) reflect an internal management structure.}
\label{exam}
\end{figure}

Network substructures -- communities -- appear in many complex networks and are frequently associated with important functions of those networks \cite{girvan2002community,spirin2003protein,krause2003compartments}. Research into  community detection has recently attracted increasing attention \cite{clauset2004finding,li2013multicomm,chakraborty2016genperm,fortunato2016community}. Meanwhile, the design of metrics to evaluate community detection has also been a recent focus of investigation \cite{newman2004finding,fortunato2007resolution,lancichinetti2008benchmark,chakraborty2017metrics}. The metrics of community structure can be generally grouped within three classes: (1) based on the density of internal links (links within the community)\cite{radicchi2004defining}; (2) based on the density of external links \cite{radicchi2004defining,fortunato2016community}; and, (3) based on the the density of both internal links and external links \cite{newman2004finding}. However, we argue that all these metrics are based on an unnecesary assumption --- that the density of internal links must be greater than that of external links. While this is true for communities in many real networks, it is not always the case. Clearly, of course, this depends on what one means by a ``community'', but we argue here that the natural intuition for community does not necessarily imply increased link density. As depicted in Fig. \ref{exam}, there are two typical communities in the natural world. Figure \ref{exam} (a) shows a mode network for which links within communities are denser than those between groups, however, Fig.\ref{exam} (b) shows another quit typical community. In this case, Fig. \ref{exam} depicts a typical work unit, and the associated subordinate relationships. The community in this case, a common work association, exhibits hierarchical rather than simply dense link structure. Moreover, the nodes group within the same region of geodesic space also can be called as the community \cite{mahmood2015subspace,mahmood2016using}. Hence, based on different kinds of real-world communities, we argue an intuitive generalisation of the existing definition of community structure. 

More importantly, the default metric Modularity \cite{newman2004finding,newman2006modularity} and its associated generalisations and extensions \cite{fortunato2007resolution,chakraborty2017metrics} are also  based on the preceding density assumption. In this paper we focus on the fact that links within a community are more {\em predictable} than external links \cite{yan2012finding,cannistraci2013link,ding2016prediction} --- due simply to the assumption that relationships within a real community are {\em stronger} (and perhaps more structured) than the external relationships. Hence, we adopt ideas from link prediction theory -- where links can be predicted based on statistical characterisation of network structure \cite{liben2007link,redner2008networks,lu2015toward,shang2019link} --- to measure the predictability of network substructure. We propose a new metric for community detection which is based on the principle that links within a community should achieve a higher link prediction performance than external links \cite{yan2012finding,cannistraci2013link,ding2016prediction}. Link prediction should work best within communities and worst between communities.

In this paper, we employ $5$ famous network open datasets to be tested. First, we use $8$ traditional community detection algorithms to uncover the communities, every algorithm has its corresponding communities for each network. Second, we analyze the link prediction performance ({\it predictability}) for the internal links and that of all links respectively. Third, we compare the performance of above two kinds of links for each network, and normalize the difference between them as the score of each detection algorithm. Meanwhile, we test the default metric {\it Modularity} for each network. We find that our metric is more stable and consistent than {\it Modularity} for each detection algorithm in all $5$ networks. Our metric can also uncover more statistical rules for each detection algorithm. We suggest that our metric not only can be more widely used, but will also achieve a more stable estimate for each algorithm. Furthermore, our metric can directly expose the failure of current community detection algorithms.

\section*{Network data}
As listed in Table~\ref{t1}, we employ $5$ open traditional network datasets that have been widely applied for community detection or link prediction problems. (1) Karate club \cite{zachary1977information}: A node represents one club user, a link indicates that there is a relationship between a pair of nodes. Karate club network is the common used data for community detection \cite{girvan2002community}. (2) Adjectives and nouns (Adjnoun)\cite{newman2006finding}: a node represents one word (adjective or noun), a link indicates that any two words are next to each other in the book. Adjectives and nouns network is also the common used data for community detection \cite{newman2006finding}. (3) Polbooks \cite{adamic2005political}: a node represents one book about US politics which is sold by Amazon.com, a link stands for there is a pair of nodes are purchased by a same buyer. Polbooks is the common used data for both community detection and link prediction \cite{adamic2005political,shang2017fitness}. (4) C.Elegans network \cite{lu2010link}: A node represents one neuron, a link indicates a synapse or a gap junction between a pair of nodes. C.Elegans network is the common used data for link prediction \cite{lu2011link,lu2010link}. (5) Usair \cite{lu2010link}: A node represents one airport, a link indicates there is a fight line between a pair of nodes. Usair network is also the common used data for link prediction \cite{lu2011link,lu2010link}.
  
\begin{table}
\caption{The number of nodes and links, and the source for $5$ undirected networks. We do not count the number of loop links and isolated nodes.}
\label{t1}
\begin{center}
\begin{tabular*}{0.43\paperwidth}{cccccc}
\hline\hline
&Karate&Adjnoun&Polbooks&C.Elegans&Usair\\ 
\hline
Nodes&$34$&$112$&$105$&$296$&$332$\\
Links&$78$&$425$&$441$&$2148$&$2126$\\ 
\hline\hline
\end{tabular*}
\end{center}
\end{table}

\section*{Methods}

\subsection*{Metric for community detection}
At the start, we use the detection algorithm to uncover the communities of network. Based on the assumption that the internal links within communities are more predictable than external or all links \cite{yan2012finding,cannistraci2013link,ding2016prediction}, we adopt three famous link prediction algorithms to test the predictable of internal links, then get the prediction accuracy scores of internal links $S_{in}^1, S_{in}^2, S_{in}^3$, and those of all links $S_{all}^1, S_{all}^2,S_{all}^3$. Then we can achieve the detection score, $$S_{pr}=\frac{\sum_{i=1}^n {\frac{S_{in}^i - S_{all}^i}{S_{all}^i}}} {n}.\eqno(1)$$ Particularly, we can adopt the link prediction algorithm for tree-like networks \cite{shang2019link} if the community structure is similar to the Figure.\ref{exam} (b).
The accuracies of link prediction algorithms for all links ($S_{all}$) will have values in $[0.5,1)$ except for that of {\it Preferential attachment index} \cite{lu2011link,shang2019link}. In addition, theoretically, $S_{in}$ should higher than $S_{all}$. Hence, for an effective community detection algorithm, the $S_{pr}$ should values in $(0,1)$ when we do not adopt the {\it Preferential Attachment} index \footnote{The AUC of the {\it Preferential attachment index} is usually lesser than $0.5$.}. Obviously, the higher value of $S_{pr}$ means the better performance of community detection algorithms. In detail, the negative value of $S_{pr}$ means the disadvantage role of the detected communities for link prediction. However, previous studies show that the community plays an advantageous role for  link prediction. That is, the corresponding detection algorithm is failing when $S_{pr}\leq0$. Generally, the effective detection algorithm should achieve a positive $S_{pr}$ value.

Moreover, we can also  extend the metric for overlapping communities. By generalising out previous arguments, we follow the principle that the scores of internal links should bigger than those of overlapping links, and the scores of overlapping links should bigger than those of external links. In this paper, we pay attention to the basic non-overlapping community problem.

\subsection*{Link prediction problem}

The graph $G$ can be completely described by a vertex set $V$, all pair of nodes set $U$, and an edge set $E$: $G=(V,E)$. The vertex set is stationary and does not evolve. Elements of the edge set are unordered pairs of elements of the vertex set: $e=(v_i,v_j)\in E$ where $v_i,v_j\in V$. The pair $(v_i,v_j)$ occurs in at most one edge $e\in E$. The edge set $E$ is divided into two parts $E^T$ and $E^P$ where $E^T\cup E^P=E$ and $E^T\cap E^P=\emptyset$. The division into $E^P$ (typically including $10\%$ of the observed links in \cite{lu2010link}) and $E^T$ (typically $90\%$ of the observed links) is arbitrary and will be used for scoring purposes. The {\em static\footnote{And this is all that we consider here.}} link prediction problem can be stated: given the training link set $E^T$ and the probe link set $E^P$ (and also $V$), $E'' = U \backslash E$, then predict a small part of unobserved links in $E''$ and that of fake links in $E^P$. That is, if we know some of the links of a network --- those links being partitioned into the {\it training set} $E^T$ and the {\it probe set} $E^P$ --- which we have {\it observed}, is it possible to predict the existence (or otherwise) of {\it unobserved} or {\it fake} links. The unobserved links are members of $E''$ and may be said to either {\it exist} or be {\it non-existent}. Generally, the link prediction scores of existence links are bigger than those of {\it fake} links or {\it non-existence} links. 

\subsubsection*{Link prediction algorithms}
The friend of our friend is our friend also, as is figured by a closed triangular structure. This common intuition is the basis of all local link prediction algorithms, with the except of the preferential attachment index. Newman et al. first use this rule to study the cooperation behaviors of scientists \cite{newman2001clustering}, which then provided a foundation for link prediction problem \cite{liben2007link}. Based on these studies, the well-known {\it common neighbors index ($CN$)} has been proposed:
$$S_{ij}^{CN}=|\Gamma(i,j)|,\eqno(2)$$
where $\Gamma(i,j)$ denotes the set of common neighbors of the nodes $i$ and $j$. The algorithm indicates that if you have a common friend with another person, there is a possible relationship between you. That is to say the friend (common neighbors) of friend is our friend.

{\it Leicht-Holme-Newman index} ($LHN1$) \cite{leicht2006vertex} directly compare the number of common neighbors and the value which is proportional to the possible number of that:
$$S_{ij}^{LHN1}= \frac{|\Gamma(i,j)|}{k(i) \times k(j)}.\eqno(3)$$

{\it Hub depressed index} ($HDI$) \cite{lu2010link} is associated with the {\it hub promoted index} ($HPI$) \cite{ravasz2002hierarchical}, both of them are based on the role of common neighbors. {\it Hub promoted index} aims at the improving of hub nodes effects. On the contrary, {\it hub depressed index} aims at the decreasing of hub nodes effects:
$$S_{ij}^{HDI}= \frac{|\Gamma(i,j)|}{max\{k(i) + k(j)\}}.\eqno(4)$$

\subsubsection*{Metric for link prediction algorithms}
The Area under Receiver Operating Characteristic Curve ($AUC$) \footnote{With the abscissa measuring the false positive rate, and the ordinate the true positive rate,  we can then draw a Receiver Operating Characteristic Curve (ROC). Statistically, the area under the ROC should be between $0.5$ and $1$. If the area is greater than $0.5$, we can suggest that our method is effective. If the area equals to $0.5$, then our method is invalid. The case that the area is less than $0.5$, is unrealistic --- in this situation the method performs so poorly that it would be better to do the reverse of what is predicted.} was originally applied to evaluate communication schemes and has since been widely applied more generally to measure prediction accuracy \cite{hanley1982meaning}. We use $AUC$ as a link prediction accuracy measure for networks. Only the information of $E^T$ is allowed to be used to compute the performance score, we compare the prediction scores of $m$ pairs of nodes from $E^P$ and $E''$ randomly, if there are $m'$ times that the score measured from $E^P$ is bigger than the score measured from $E''$ and $m''$ times that the two scores are equal, then, the prediction accuracy $AUC=(m'+0.5m'')/m$. In this letter, we compute the AUC of all algorithms $100$ times independently.

\section*{Results}
Similar to previous studies \cite{girvan2002community}, we focus on correctly identifying links within communities. Assuming that links within a community are more predictable than external links, we compare the accuracy of link prediction for internal links and that of all (internal and external) links. Here, we adopt $8$ traditional community detection methods to be tested. (1) Kclique algorithm is based on the relationship between nodes and subgraphs \cite{palla2005uncovering}; (2) Fast greedy (FG) adopts the greedy optimization method \cite{clauset2004finding};  (3) Girvan-Newman algorithm (GN) is based on the principle of edge betweenness \cite{girvan2002community}; (4) Label Propagation algorithm (LP) is based on the propagation process of labeled nodes \cite{xie2012towards}; (5) the leading eigenvector of modularity matrix is the key for Leading eigenvector algorithm (LE) \cite{newman2006modularity}; (6) MultiLevel algorithm actually is the multi-resolution version of modularity \cite{breiger1975algorithm}; (7) WalkTrap algorithm adopts the random walk method \cite{pons2005computing}; and (8) InfoMap algorithm is based on random walk dynamics \cite{rosvall2008maps}. These diverse algorithms make it more accurate for analyzing the predictability of community structures. 

\subsection*{Internal links versus all links}
\begin{figure}[htbp]
\centering
\includegraphics[width=\linewidth]{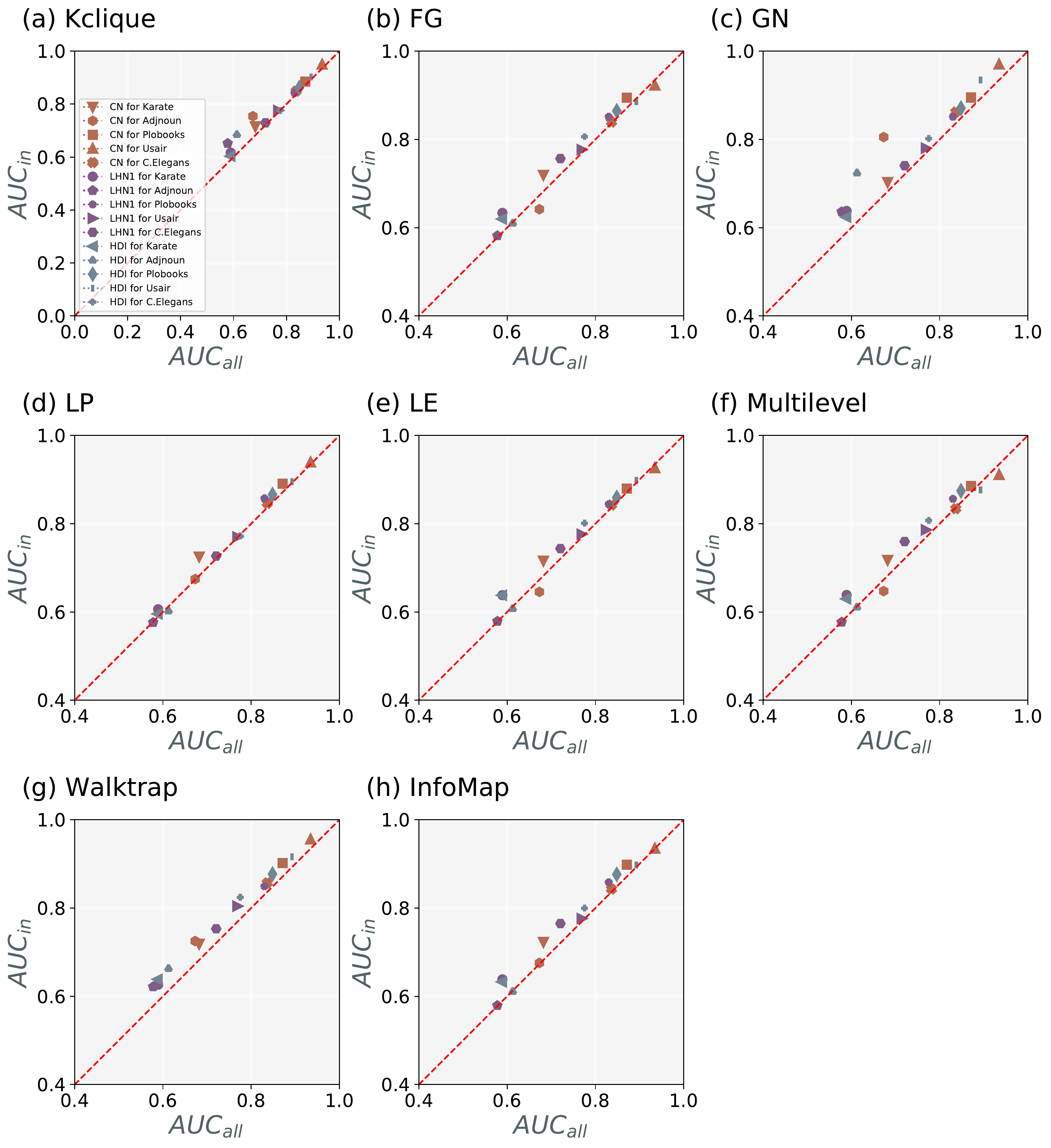}
\caption{For $8$ traditional community detection algorithms, the prediction accuracies of community internal links versus all links via three traditional link prediction algorithms in $5$ traditional networks. The ordinate is the AUC score of internal links, and the abscissa is the AUC score of all links. The dashed line is the diagonal. Obviously, the dot which is above the diagonal means the AUC score of internal links is bigger than that of all links for the corresponding link prediction algorithm.}
\label{invall}
\end{figure}

As shown in Figure.~\ref{invall}, we can see that, for almost all link prediction algorithms, the prediction accuracies of internal links is higher than that of all links. That is to say, the internal links are more predictable than all links, further than external links. This result also provides the design basis for our community detection metric. We also observe that for the detection algorithms {\it FG}, {\it LE} and {\it Multilevel}, the {\it CN} accuracies of internal links are obviously lower than those of all links in the Adjnoun network and the Usair network.

\begin{figure}[htbp]
\centering
\includegraphics[width=\linewidth]{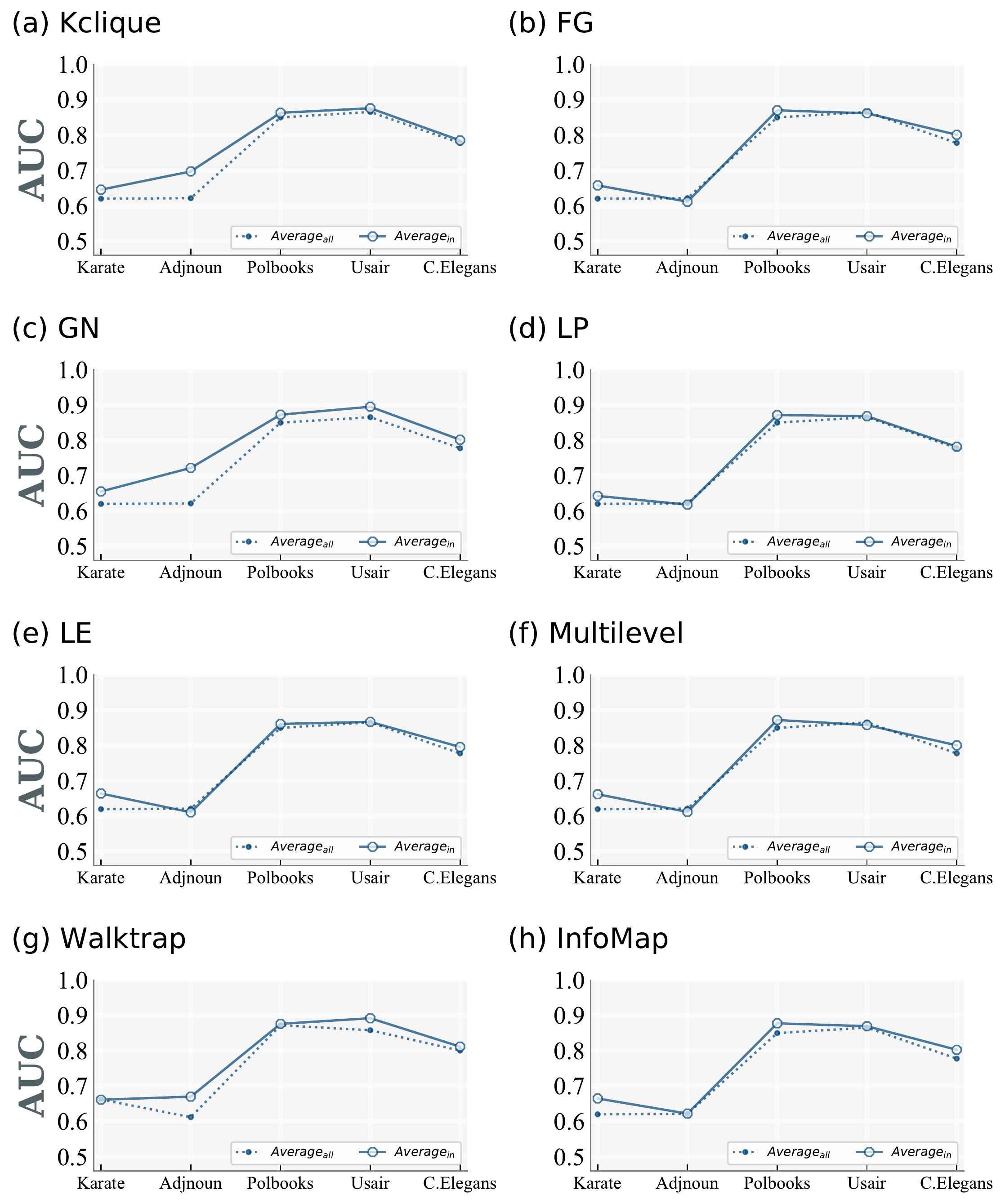}
\caption{For $8$ traditional community detection algorithms, the average prediction accuracies of community internal links and all links of three traditional link prediction algorithms in $5$ traditional networks. The hollow dots indicate the results of internal links, and the filled dots indicate the results of all links.}
\label{ave}
\end{figure}

Hence, to clearly distinguish the qualities of detection algorithms. As depicted in Figure.~\ref{ave}, we show the average link prediction accuracies of all three link prediction algorithms for each community detection algorithm. We can clearly see that, in general, the prediction accuracies of internal links are higher than those of all links. In detail, for the algorithms {\it Kclique}, {\it Girvan-Newman} and {\it WalkTrap}, the prediction accuracies of internal links always are higher than that of all links. On the contrary, for algorithms {\it FG}, {\it LE} and {\it Multilevel}, the prediction accuracies of internal links are lower than those of all links in the Adjnoun network and the Usair network, namely these community detection algorithms are not effective for the Adjnoun network and the Usair network.

\subsection*{Predictability versus Modularity}
\begin{figure}[htbp]
\centering
\includegraphics[width=\linewidth]{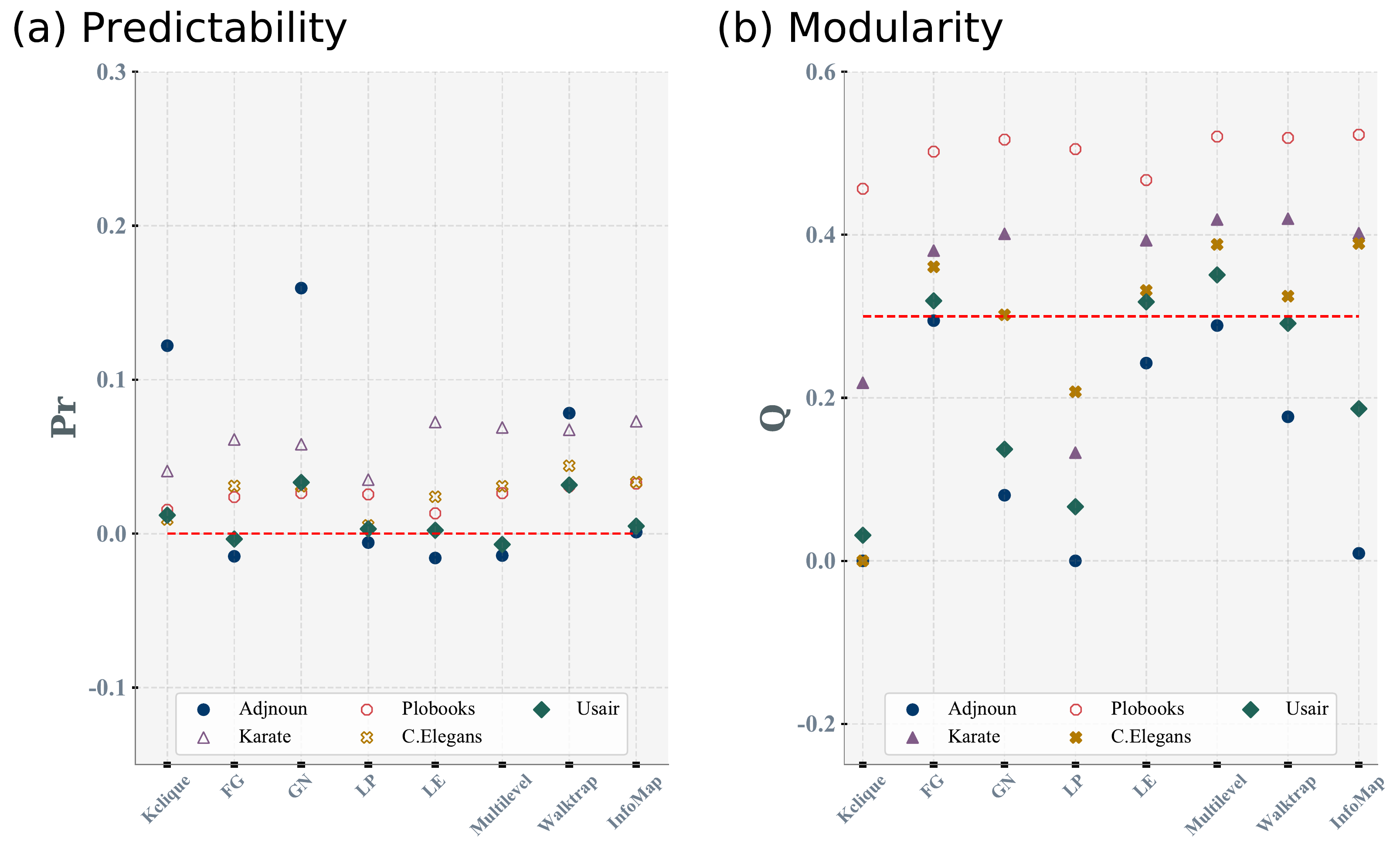}
\caption{For $5$ famous networks, the results of metric {\it Predictability} and metric {\it Modularity}. The ordinate is the score of corresponding metric. The dashed line indicates the effective score. The hollow marks indicate the results of corresponding networks are more stable than those of other networks.}
\label{pr}
\end{figure}

In what follows, we directly compare our metric and the default metric {\it Modularity} \cite{newman2004finding,newman2006modularity}. As shown in Figure.~\ref{pr} (a), under our metric, we can see that the same community detection algorithm performs similar in different networks. On the contrary, under the default metric {\it Modularity}, the performance of the same community detection algorithm varies greatly from network to network (Figure.~\ref{pr} (b)). Hence, as shown in Table.~\ref{t2}, we adopt the standard deviation to further measure the robustness for these two metrics. The standard deviation of our metric is much less than that of {\it Modularity}. Obviously, our metric {\it Predictability} is more stable than {\it Modularity}. Hence, we infer that the {\it Modularity} cannot consistently measure a community detection algorithm in different networks. Our metric does lead to a consistent  and stable evaluation for the same community detection algorithm between different networks. 

\begin{table*}
\caption{For all networks, the standard deviations of {\it Predictability} metric and {\it Modularity} metric.}
\label{t2}
\begin{center}
\begin{tabular*}{0.53\paperwidth}{cccccc}
\hline\hline
Standard deviation&Karate&Adjnoun&Polbooks&C.Elegans&Usair\\ 
\hline
{\it Predictability}&$0.0135$&$0.0663$&$0.0063$&$0.0120$&$0.0142$\\
{\it Modularity}&$0.1012$&$0.1212$&$0.2389$&$0.1217$&$0.1163$\\ 
\hline\hline
\end{tabular*}
\end{center}
\end{table*}

In addition, our metric can directly and carefully judge the failure of community detection algorithms -- if a community detection algorithm achieves a negative {\it Predictability} score, we can infer the failure of community detection. For example, as depicted in Figure.~\ref{pr} (a), for the Adjnoun network and the Usair network, the $Fast greedy$ algorithm and $Multilevel$ algorithm both achieve the negetive scores, which means that they cannot divide the communities successfully.

More importantly, we also find that the algorithms {\it Kclique}, {\it Girvan-Newman} and {\it WalkTrap} are successful for each network (Figure.~\ref{pr} (a)). Furthermore, {\it Modularity} can indirectly measure the ( $Q \geq 0.3$\cite{clauset2004finding}) the failure of community detection. However, as shown in Figure.~\ref{pr} (b), we cannot find any successful community detection algorithm for all networks. We suggest that the weak robustness of ${\it Modularity}$ is the key reason for that phenomenon. And our metric {\it Predictability} can capture more common statistical characteristics for the community detection algorithms. 

As shown in Figure.~\ref{pr} (b), we can find that the results of the Polbooks network are more stable than other networks, which means that ${\it Modularity}$ also can capture the common statistical characteristic for the community detection algorithm. However, as shown in Figure.~\ref{pr} (a), we not only find the results of Polbooks network are stable and always bigger than $0$, but also observe that both the results of Karate network and C.Elegans network are stable and always bigger than $0$. That phenomenon further suggest that our metric can capture more common statistical rules for community detection algorithms. Our metric can find more networks that can easy be divided by the community algorithm. 

\section*{Conclusion and Discussion}

Though there are many competing  metrics with which to evaluate community detection, the different statistical characteristics between internal links of communities and external links of communities is always the key principle. In this paper, we argue that the principle that the density of internal links is higher than that of external links, is not a universal characteristic of all ``communities'' for all real-world networks. Hence, we propose the {\it Predictability} metric which is based on the principle that internal links are more predictable than external links. We find that our metric can reliably evaluate the performance of the same community detection algorithms across a range of different networks, however, the default metric {\it Modularity} is rather unstable. We infer that the unstable performance of {\it Modularity} is due to the ill-defined of communities: the density of links within communities is higher than that of external links.

Furthermore, our metric can evaluate the performance of community detection algorithm more carefully -- the failure of community detection can be uncovered and more statistical rules can be found by our metric. In detail, the negative score of {\it Predictability} directly reflects the bad or failure level of a community detection algorithm, inversely, the positive score of {\it Predictability} reflects the good level of a community detection algorithm.  We also can find more networks that can be divided by the community algorithm.
We suggest that the different predictabilities between internal links and external links provide more information than the traditional principle of community detection metrics, our metric can help researchers further discuss the precise definition of the community detection problem.

\bibliographystyle{aaai}


\begin{thebibliography}{}

\bibitem[\protect\citeauthoryear{Adamic and Glance}{2005}]{adamic2005political}
Adamic, L.~A., and Glance, N.
\newblock 2005.
\newblock The political blogosphere and the 2004 us election: divided they
  blog.
\newblock In {\em Proceedings of the 3rd International Workshop on Link
  Discovery},  36--43.
\newblock ACM.

\bibitem[\protect\citeauthoryear{Breiger, Boorman, and
  Arabie}{1975}]{breiger1975algorithm}
Breiger, R.~L.; Boorman, S.~A.; and Arabie, P.
\newblock 1975.
\newblock An algorithm for clustering relational data with applications to
  social network analysis and comparison with multidimensional scaling.
\newblock {\em Journal of Mathematical Psychology} 12(3):328--383.

\bibitem[\protect\citeauthoryear{Cannistraci, Alanis-Lobato, and
  Ravasi}{2013}]{cannistraci2013link}
Cannistraci, C.~V.; Alanis-Lobato, G.; and Ravasi, T.
\newblock 2013.
\newblock From link-prediction in brain connectomes and protein interactomes to
  the local-community-paradigm in complex networks.
\newblock {\em Scientific Reports} 3:1613.

\bibitem[\protect\citeauthoryear{Chakraborty \bgroup et al\mbox.\egroup
  }{2016}]{chakraborty2016genperm}
Chakraborty, T.; Kumar, S.; Ganguly, N.; Mukherjee, A.; and Bhowmick, S.
\newblock 2016.
\newblock Genperm: a unified method for detecting non-overlapping and
  overlapping communities.
\newblock {\em IEEE Transactions on Knowledge and Data Engineering}
  28(8):2101--2114.

\bibitem[\protect\citeauthoryear{Chakraborty \bgroup et al\mbox.\egroup
  }{2017}]{chakraborty2017metrics}
Chakraborty, T.; Dalmia, A.; Mukherjee, A.; and Ganguly, N.
\newblock 2017.
\newblock Metrics for community analysis: A survey.
\newblock {\em ACM Computing Surveys (CSUR)} 50(4):54.

\bibitem[\protect\citeauthoryear{Clauset, Newman, and
  Moore}{2004}]{clauset2004finding}
Clauset, A.; Newman, M.~E.; and Moore, C.
\newblock 2004.
\newblock Finding community structure in very large networks.
\newblock {\em Physical Review E} 70(6):066111.

\bibitem[\protect\citeauthoryear{Ding \bgroup et al\mbox.\egroup
  }{2016}]{ding2016prediction}
Ding, J.; Jiao, L.; Wu, J.; and Liu, F.
\newblock 2016.
\newblock Prediction of missing links based on community relevance and ruler
  inference.
\newblock {\em Knowledge-Based Systems} 98:200--215.

\bibitem[\protect\citeauthoryear{Fortunato and
  Barthelemy}{2007}]{fortunato2007resolution}
Fortunato, S., and Barthelemy, M.
\newblock 2007.
\newblock Resolution limit in community detection.
\newblock {\em Proceedings of the National Academy of Sciences} 104(1):36--41.

\bibitem[\protect\citeauthoryear{Fortunato and
  Hric}{2016}]{fortunato2016community}
Fortunato, S., and Hric, D.
\newblock 2016.
\newblock Community detection in networks: A user guide.
\newblock {\em Physics Reports} 659:1--44.

\bibitem[\protect\citeauthoryear{Girvan and Newman}{2002}]{girvan2002community}
Girvan, M., and Newman, M.~E.
\newblock 2002.
\newblock Community structure in social and biological networks.
\newblock {\em Proceedings of the National Academy of Sciences}
  99(12):7821--7826.

\bibitem[\protect\citeauthoryear{Hanley and McNeil}{1982}]{hanley1982meaning}
Hanley, J.~A., and McNeil, B.~J.
\newblock 1982.
\newblock The meaning and use of the area under a receiver operating
  characteristic (roc) curve.
\newblock {\em Radiology} 143(1).

\bibitem[\protect\citeauthoryear{Krause \bgroup et al\mbox.\egroup
  }{2003}]{krause2003compartments}
Krause, A.~E.; Frank, K.~A.; Mason, D.~M.; Ulanowicz, R.~E.; and Taylor, W.~W.
\newblock 2003.
\newblock Compartments revealed in food-web structure.
\newblock {\em Nature} 426(6964):282.

\bibitem[\protect\citeauthoryear{Lancichinetti, Fortunato, and
  Radicchi}{2008}]{lancichinetti2008benchmark}
Lancichinetti, A.; Fortunato, S.; and Radicchi, F.
\newblock 2008.
\newblock Benchmark graphs for testing community detection algorithms.
\newblock {\em Physical Review E} 78(4):046110.

\bibitem[\protect\citeauthoryear{Leicht, Holme, and
  Newman}{2006}]{leicht2006vertex}
Leicht, E.~A.; Holme, P.; and Newman, M.~E.
\newblock 2006.
\newblock Vertex similarity in networks.
\newblock {\em Physical Review E} 73(2):026120.

\bibitem[\protect\citeauthoryear{Li, Ng, and Ye}{2013}]{li2013multicomm}
Li, X.; Ng, M.~K.; and Ye, Y.
\newblock 2013.
\newblock Multicomm: Finding community structure in multi-dimensional networks.
\newblock {\em IEEE Transactions on Knowledge and Data Engineering}
  26(4):929--941.

\bibitem[\protect\citeauthoryear{Liben-Nowell and
  Kleinberg}{2007}]{liben2007link}
Liben-Nowell, D., and Kleinberg, J.
\newblock 2007.
\newblock The link-prediction problem for social networks.
\newblock {\em Journal of the American Society for Information Science and
  Technology} 58(7):1019--1031.

\bibitem[\protect\citeauthoryear{L{\"u} and Zhou}{2010}]{lu2010link}
L{\"u}, L., and Zhou, T.
\newblock 2010.
\newblock Link prediction in weighted networks: The role of weak ties.
\newblock {\em EPL} 89(1):18001.

\bibitem[\protect\citeauthoryear{L{\"u} and Zhou}{2011}]{lu2011link}
L{\"u}, L., and Zhou, T.
\newblock 2011.
\newblock Link prediction in complex networks: A survey.
\newblock {\em Physica A: Statistical Mechanics and its Applications}
  390(6):1150--1170.

\bibitem[\protect\citeauthoryear{L{\"u} \bgroup et al\mbox.\egroup
  }{2015}]{lu2015toward}
L{\"u}, L.; Pan, L.; Zhou, T.; Zhang, Y.-C.; and Stanley, H.~E.
\newblock 2015.
\newblock Toward link predictability of complex networks.
\newblock {\em Proceedings of the National Academy of Sciences}
  112(8):2325--2330.

\bibitem[\protect\citeauthoryear{Mahmood and Small}{2015}]{mahmood2015subspace}
Mahmood, A., and Small, M.
\newblock 2015.
\newblock Subspace based network community detection using sparse linear
  coding.
\newblock {\em IEEE Transactions on Knowledge and Data Engineering}
  28(3):801--812.

\bibitem[\protect\citeauthoryear{Mahmood \bgroup et al\mbox.\egroup
  }{2016}]{mahmood2016using}
Mahmood, A.; Small, M.; Al-Maadeed, S.~A.; and Rajpoot, N.
\newblock 2016.
\newblock Using geodesic space density gradients for network community
  detection.
\newblock {\em IEEE Transactions on Knowledge and Data Engineering}
  29(4):921--935.

\bibitem[\protect\citeauthoryear{Newman and Girvan}{2004}]{newman2004finding}
Newman, M.~E., and Girvan, M.
\newblock 2004.
\newblock Finding and evaluating community structure in networks.
\newblock {\em Physical Review E} 69(2):026113.

\bibitem[\protect\citeauthoryear{Newman}{2001}]{newman2001clustering}
Newman, M.~E.
\newblock 2001.
\newblock Clustering and preferential attachment in growing networks.
\newblock {\em Physical Review E} 64(2):025102.

\bibitem[\protect\citeauthoryear{Newman}{2006a}]{newman2006finding}
Newman, M.~E.
\newblock 2006a.
\newblock Finding community structure in networks using the eigenvectors of
  matrices.
\newblock {\em Physical Review E} 74(3):036104.

\bibitem[\protect\citeauthoryear{Newman}{2006b}]{newman2006modularity}
Newman, M.~E.
\newblock 2006b.
\newblock Modularity and community structure in networks.
\newblock {\em Proceedings of the National Academy of Sciences}
  103(23):8577--8582.

\bibitem[\protect\citeauthoryear{Palla \bgroup et al\mbox.\egroup
  }{2005}]{palla2005uncovering}
Palla, G.; Der{\'e}nyi, I.; Farkas, I.; and Vicsek, T.
\newblock 2005.
\newblock Uncovering the overlapping community structure of complex networks in
  nature and society.
\newblock {\em Nature} 435(7043):814.

\bibitem[\protect\citeauthoryear{Pons and Latapy}{2005}]{pons2005computing}
Pons, P., and Latapy, M.
\newblock 2005.
\newblock Computing communities in large networks using random walks.
\newblock In {\em International Symposium on Computer and Information
  Sciences},  284--293.
\newblock Springer.

\bibitem[\protect\citeauthoryear{Radicchi \bgroup et al\mbox.\egroup
  }{2004}]{radicchi2004defining}
Radicchi, F.; Castellano, C.; Cecconi, F.; Loreto, V.; and Parisi, D.
\newblock 2004.
\newblock Defining and identifying communities in networks.
\newblock {\em Proceedings of the National Academy of Sciences}
  101(9):2658--2663.

\bibitem[\protect\citeauthoryear{Ravasz \bgroup et al\mbox.\egroup
  }{2002}]{ravasz2002hierarchical}
Ravasz, E.; Somera, A.~L.; Mongru, D.~A.; Oltvai, Z.~N.; and Barab{\'a}si,
  A.-L.
\newblock 2002.
\newblock Hierarchical organization of modularity in metabolic networks.
\newblock {\em Science} 297(5586):1551--1555.

\bibitem[\protect\citeauthoryear{Redner}{2008}]{redner2008networks}
Redner, S.
\newblock 2008.
\newblock Networks: teasing out the missing links.
\newblock {\em Nature} 453(7191):47--48.

\bibitem[\protect\citeauthoryear{Rosvall and Bergstrom}{2008}]{rosvall2008maps}
Rosvall, M., and Bergstrom, C.~T.
\newblock 2008.
\newblock Maps of random walks on complex networks reveal community structure.
\newblock {\em Proceedings of the National Academy of Sciences}
  105(4):1118--1123.

\bibitem[\protect\citeauthoryear{Shang \bgroup et al\mbox.\egroup
  }{2019}]{shang2019link}
Shang, K.-k.; Li, T.-c.; Small, M.; Burton, D.; and Wang, Y.
\newblock 2019.
\newblock Link prediction for tree-like networks.
\newblock {\em Chaos: An Interdisciplinary Journal of Nonlinear Science}
  29(6):061103.

\bibitem[\protect\citeauthoryear{Shang, Small, and
  Yan}{2017}]{shang2017fitness}
Shang, K.-k.; Small, M.; and Yan, W.-s.
\newblock 2017.
\newblock Fitness networks for real world systems via modified preferential
  attachment.
\newblock {\em Physica A: Statistical Mechanics and its Applications}
  474:49--60.

\bibitem[\protect\citeauthoryear{Spirin and Mirny}{2003}]{spirin2003protein}
Spirin, V., and Mirny, L.~A.
\newblock 2003.
\newblock Protein complexes and functional modules in molecular networks.
\newblock {\em Proceedings of the National Academy of Sciences}
  100(21):12123--12128.

\bibitem[\protect\citeauthoryear{Xie and Szymanski}{2012}]{xie2012towards}
Xie, J., and Szymanski, B.~K.
\newblock 2012.
\newblock Towards linear time overlapping community detection in social
  networks.
\newblock In {\em Pacific-Asia Conference on Knowledge Discovery and Data
  Mining},  25--36.
\newblock Springer.

\bibitem[\protect\citeauthoryear{Yan and Gregory}{2012}]{yan2012finding}
Yan, B., and Gregory, S.
\newblock 2012.
\newblock Finding missing edges in networks based on their community structure.
\newblock {\em Physical Review E} 85(5):056112.

\bibitem[\protect\citeauthoryear{Zachary}{1977}]{zachary1977information}
Zachary, W.~W.
\newblock 1977.
\newblock An information flow model for conflict and fission in small groups.
\newblock {\em Journal of Anthropological Research} 33(4):452--473.

\end{thebibliography}

\section*{Acknowledgements}
Ke-ke Shang is supported by National Natural Science Foundation of China 61803047 and Tencent Research Institute. Michael Small is supported by ARC Discovery Project DP180100718.

\section*{Author contributions statement}
Ke-ke Shang conceived the experiments,  Yan Wang conducted the experiments, Ke-ke Shang, Michael Small, Yan Wang and Di Yin analyzed the results, Ke-ke Shang, Michael Small and Shu Li wrote the manuscript. All authors reviewed the manuscript.

\section*{Competing interests} 
The authors declare no competing interests.

\section*{Data Availability}
The datasets generated and/or analysed in this study are available from the corresponding author on reasonable request.


\end{document}